\documentstyle[prl,aps,preprint,eqsecnum]{revtex}
\begin{document}
\draft
\title{Massless scalar fields and topological black holes}

\author{Tekin Dereli and Yuri N.\ Obukhov\footnote{On leave from: 
Department of Theoretical Physics, Moscow State University, 
117234 Moscow, Russia}}
\address
{Department of Physics, 
Middle East Technical University, 06531 Ankara, Turkey}

\maketitle

\bigskip
\noindent{\it file scalar.tex, 24 September 1999, draft}
\bigskip

\begin{abstract}
The exact static solutions in the higher dimensional Einstein-Maxwell-Klein-Gordon 
theory are investigated. With the help of the methods developed for the effective 
dilaton type gauge gravity models in two dimensions, we find new spherically
 and hyperbolically symmetric solutions which generalize the four dimensional 
configurations of Dereli-Eri\c{s}. We show that, like in four dimensions, the 
non-trivial scalar field yields, in general, a naked singularity. The new 
solutions are compared with the higher dimensional Brans-Dicke black hole type
solutions. 
\end{abstract}

\section{Introduction}

The study of exact solutions remains one of the central issues of Einstein's 
gravitational theory and its generalizations. Both the higher-dimensional and 
lower-dimensional gravity models (see, e.g., \cite{highlow}) have attracted 
considerable attention recently. This interest was partly motivated by the 
idea that the geometrical structures not confined to the dimension four may 
be helpful in understanding the four-dimensional physics, and partly was a
due to the development of (super)unification approaches. 

Topological black holes possess horizons with non-trivial topology. They 
naturally arise in a number of gravitational models \cite{manntop,cai1,lemos} 
(in particular, as black strings, black branes, etc) and have rather different
properties from the usual black holes with spherical horizons. Higher 
dimensional generalizations of these object were considered recently in
\cite{cai2}. 

As it is well known, the non-trivial scalar field may, in general, destroy a
horizon and produce a naked singularity instead. This fact, first noticed
in four dimensions, cf. \cite{jan}, was later confirmed also for the higher
dimensions \cite{xan,agnese}, although for certain types of dilaton couplings
the higher-dimensional black holes with scalar fields do exist \cite{GM,kim}.
For a comprehensive recent review see \cite{youm}. 

It seems interesting to consider Einstein's gravity with all the three above 
mentioned features combined: i.e. in an arbitrary dimensional spacetime, with 
a scalar field, and possibly with non-trivial topology of horizons. This problem 
is studied in the present paper. Technically, we use the methods developed
previously for the two-dimensional gauge gravity models \cite{2drc,tworev}.
For this purpose, in Sect.~\ref{KK}, the original problem is reduced from
$(d+n)$ dimensions to an effective $d$-dimensional model by means of the 
Kaluza-Klein scheme with an ``internal'' $n$-dimensional space of constant
curvature (positive, negative or zero). Then, in Sect.~\ref{2D}, the case $d=2$ 
is analyzed in detail with the help of the technique developed in \cite{2drc,tworev}.
The new exact solutions are derived in Sect.~\ref{sol}, and we discuss their
properties and make conclusions in Sect.~\ref{disc}.

\section{Kaluza-Klein scheme}\label{KK}

Let us consider the Kaluza-Klein reduction of a $(d+n)$-dimensional manifold
to the physical $d$-dimensional Riemannian spacetime $M^d$ with an $n$-dimensional
internal space of constant curvature. Denote the components of the higher
dimensional curvature two-form ${\cal R}^{AB}$ with respect to a local orthonormal
frame $E_A$. The dual coframe one-forms are denoted $\vartheta^A$, and the
indices run $A,B,...=0,1,\dots,d+n-1$. The general Kaluza-Klein reduction of an
$(d+n)$-dimensional manifold with a compactification on an $n$-torus involves
$n$ 1-forms $A^a$ and $n^2$ scalar fields $\Phi^a_b$. Here we will consider
a simplified scheme without the Kaluza-Klein 1-forms (gauge fields). Then the
consistent decomposition of the metric reads:
\begin{equation}
{\buildrel (d+n)\over g}\,=\,{\buildrel (d)\over g}\, +\, e^{-{4\over n}\Phi}\,
{\buildrel (n)\over g}.\label{KK0}
\end{equation}
Here $\Phi$ is the Kaluza-Klein scalar field (only one scalar survives in absence
of $A^a$) which depends only on the coordinates of $M^d$, and 
\begin{eqnarray}
{\buildrel (d)\over g}&=& g_{\alpha\beta}\,\vartheta^\alpha\otimes\vartheta^\beta,\\
{\buildrel (n)\over g}&=& g_{ab}\,\vartheta^a\otimes\vartheta^b,
\end{eqnarray}
describe, respectively, the metric of the physical spacetime [with $g_{\alpha\beta}=
{\rm diag}(-1,1,\dots,1)$ as a $d$-dimensional Minkowski metric] and the internal 
space [with $g_{ab}=\delta_{ab}$] of a constant curvature 
$R^{ab}=-\,\lambda\,\vartheta^a\wedge\vartheta^b$. The constant
$\lambda=+1$ for an $n$-sphere of a unit radius, $\lambda=0$ for flat space 
(e.g., hyperplane, cylinder or $n$-torus), and $\lambda=-1$ for a hyperbolic space.

The (local frame) indices clearly run: $\alpha,\beta,\dots = 0,1,\dots,d-1$, and 
$a,b,\dots =1,\dots,n$.

The Einstein-Maxwell-Klein-Gordon theory with a cosmological term 
in $d+n$ dimensions reads
\begin{equation}
L = -\,{1\over 2}\,{\cal R}^{AB}\wedge\eta_{AB} - {1\over 2}\,F\wedge
\hbox{$\scriptstyle{\#}$} F -\,{1\over 2}\left(d\phi\wedge\hbox{$\scriptstyle{\#}$} 
d\phi + m^2\,\phi\hbox{$\scriptstyle{\#}$}\phi\right) -\Lambda\eta.\label{Lhigh}
\end{equation}
Here $F=dA$ is the Maxwell field strength two-form and $\phi$ is the scalar field.
We are using the general notations and conventions of \cite{PR}. In particular,
the Trautman's $\eta$-basis of exterior forms is defined by the Hodge duals 
of the products of coframe one-forms $\vartheta^A$: given the volume $(d+n)$-form 
$\eta$, one has $\eta_A = \hbox{$\scriptstyle{\#}$}\vartheta_A = E_A\rfloor\eta$,
$\eta_{AB} = \hbox{$\scriptstyle{\#}$}(\vartheta_A\wedge\vartheta_B) = 
E_A\rfloor\eta_B$, etc. Same notation is used for the lower-dimensional
counterparts in $M^d$.

We will consider a massless Klein-Gordon field, so that $m^2=0$. 

Assuming that the Maxwell and scalar fields are independent of the internal
space coordinates, we straightforwardly obtain from (\ref{Lhigh}) a dimensionally
reduced Lagrangian:
\begin{eqnarray}
L &=& e^{-2\Phi}\Bigg(-\,{1\over 2}\,R^{\alpha\beta}\wedge\eta_{\alpha\beta} 
+ 2\,{n-1\over n}\,d\Phi\wedge\ast d\Phi + \,{1\over 2}\,{\buildrel (n)\over R}\,
e^{{4\over n}\Phi}\,\eta \nonumber\\ &&\qquad\qquad -\,{1\over 2}\,F\wedge\ast F -
\,{1\over 2}\,d\phi\wedge\ast d\phi- \Lambda\eta \Bigg).\label{Lred}
\end{eqnarray}
Here: ${\buildrel (n)\over R}=\lambda\,n(n-1)$ is the curvature scalar of the
internal space, and from now on $\eta$ denotes the volume $d$-form and $\ast$ is
the $d$-dimensional Hodge operator on $M^d$. 

It seems worthwhile to note that Kaluza-Klein reduction from $(d + n)$ 
dimensions in the limit of $n\rightarrow\infty$ yields exactly the low-energy 
string model in an arbitrary dimension $d$ {\cite{CQG}}. 

\section{Case of $d=2$: effective two-dimensional theory}\label{2D}

Let us put $d=2$. The above compactification obviously describes the general
$2+n$-dimensional metric configurations with spherical ($\lambda=1$), plane 
($\lambda=0$) and ``hyperboloidal'' ($\lambda=-1$) symmetry. The reduced system 
(\ref{Lred}) gives the dynamics of the ``radial'' variables in terms of a
dilaton type gravity theory in two dimensions. 

Recently there was an increasing interest in the so-called topological black holes 
which are defined as solutions of Einstein field equations for $\lambda=0,-1$.

We will look for the exact solutions of the Einstein-Maxwell-scalar field
equations using the methods developed for the general two-dimensional
Poincar\'e gauge gravity \cite{2drc}, see the review \cite{tworev} which 
contains also a list of references to other approaches. In particular, as it 
is demonstrated in \cite{tworev}, any two-dimensional dilaton type model
with a Lagrangian $\eta\left[{\cal F}(\Phi)\widetilde{R} + {\cal G}(\Phi)
(\partial_\alpha\Phi)^2 + {\cal U}(\Phi)\right]$ can be recasted into a form 
of an effective theory in the Riemann-Cartan spacetime, and the gravitational 
field equations then can be straightforwardly integrated. The key role in this
approach plays a two-dimensional torsion trace one-form $T$ which, together 
with the Hodge dual $\ast T$, provides a natural coframe basis in the spacetime. 

For the Lagrangian (\ref{Lred}) we have
\begin{equation}
{\cal F}(\Phi) = {1\over 2}\,e^{-2\Phi},\qquad {\cal G}(\Phi) = 2\left(
{n-1\over n}\right)e^{-2\Phi},\qquad {\cal U}(\Phi) = {\lambda\over 2}\,n(n-1)
\,e^{-2\left({n-2\over n}\right)\Phi}.\label{FGU}
\end{equation}
Correspondingly, one finds the Lagrangian of the equivalent Einstein-Cartan theory:
\begin{equation}
L_{\rm EC} = -\,{1\over 2}\,\xi(\Phi)\,T^\alpha\wedge\ast T^\alpha 
-{1\over 2}\,\omega(\Phi)\,R^{\alpha\beta}\,\eta_{\alpha\beta}\, +\,
{\cal U}(\Phi)\,\eta,\label{Leff}
\end{equation}
where 
\begin{equation}
\omega = e^{-2\Phi},\qquad\qquad\xi=\left({n\over n-1}\right)e^{-2\Phi}.\label{omxi}
\end{equation}
Here $T^\alpha$ is the torsion two-form and the curvature is also constructed
from the Lorentz connection with torsion. In two dimensions, the Hodge dual
$t^\alpha = \ast T^\alpha$ is a (covector-valued) scalar. The torsion trace
one-form $T:=e_\alpha\rfloor T^\alpha = - t^\alpha\eta_\alpha$ together with 
$\ast T$ form a basis of the cotangent space when $t^2:=t^\alpha t_\alpha$ is
non-zero. As a consequence, the two-metric arises as 
\begin{equation}
{\buildrel (2) \over g} = - \left(\vartheta^0\right)^2 + \left(
\vartheta^1\right)^2 = {1\over -t^2}\left[(T)^2 - (\ast T)^2\right].\label{met1}
\end{equation}
[Note that $t^2$ is a negative quantity]. The gravitational field equations, which
arise from the variations of the action with respect to the coframe and connection
one-forms (Palatini principle), after some rearrangements can be written as
\begin{eqnarray}
d(\xi^2 t^2) &=& \left(\xi^2 t^2 - 2\xi{\cal U}\right)T + 2\xi\,S,\label{eq1}\\
d(\xi\ast\! T) &=& 2{\cal U}\,\eta + \vartheta^\alpha\wedge\Sigma_\alpha,\label{eq2}\\
-\,2\left({n-1\over n}\right)\,d\Phi &=& T.\label{eq3}
\end{eqnarray}
Here $S:=t^\alpha\Sigma_\alpha$, and as usual the source is represented by the
energy-momentum one form which is obtained as a variational derivative of the
matter Lagrangian [second line in (\ref{Lred})] with respect to the coframe: 
\begin{equation}
\Sigma_\alpha = -\,e^{-2\Phi}\Lambda\eta_\alpha + {1\over 2}\,e^{-2\Phi}\,
(e_\alpha\rfloor F)\ast\! F  +\,{1\over 2}\,e^{- 2\Phi}\left[(e_a\rfloor
d\phi)\ast\! d\phi + d\phi (e_a\rfloor\!\ast\! d\phi)\right].\label{mom}
\end{equation}
From (\ref{met1}) and (\ref{eq3}) we conclude that $\Phi$ can be taken as a local 
{\it spatial} coordinate, and one can construct a second leg of the coframe as 
\begin{equation}
\xi\ast\! T = Bd\tau,\label{TB} 
\end{equation}
where $\tau$ is the local {\it time} coordinate, and $B=B(\tau,\Phi)$ is some yet 
unknown function.

The Maxwell equation $d\left(e^{-2\Phi}\ast\! F\right)=0$ and the Klein-Gordon
equation $d\left(e^{-2\Phi}\ast\!d\phi\right)=0$ are easily integrated, yielding
\begin{eqnarray}
\ast F &=& Q\,e^{2\Phi},\label{Qc1}\\
B\,\phi' &=& c_0, \label{Qc2}
\end{eqnarray} 
where $Q$ and $c_0$ are integration constants. Hereafter the prime denotes the
derivative with respect to $\Phi$. 

Substituting (\ref{TB}) and (\ref{Qc1})-(\ref{Qc2}) into (\ref{mom}), and 
subsequently into (\ref{eq1})-(\ref{eq2}), one finds a system of two differential 
equations for the unknown functions $(\xi^2 t^2)$ and $B$:
\begin{eqnarray}
(\xi^2 t^2)' &=& -\,2\left({n-1\over n}\right)\xi^2 t^2 
+ 2\lambda\,n(n-1)\,e^{-4\left({n-1\over n}\right)\Phi} - 4\Lambda e^{-4\Phi} 
- 2Q^2 + {c_0^2\over 2}\,{\xi^2 t^2\over B^2},\label{XT0}\\
{(B^2)'\over 2B^2} &=& {2\lambda\,n(n-1)\,e^{-4\left({n-1\over n}\right)\Phi} 
- 4\Lambda e^{-4\Phi} - 2Q^2 \over \xi^2 t^2}.\label{B0}
\end{eqnarray}
It seems worthwhile to note that although most of the intermediate derivations
were, strictly speaking, inapplicable to the case of $n=1$, the final system
is meaningful also when $n=1$, yielding correct solutions. 

Consider at first the case $n=2$ {\it and} $\lambda=1$. Then $\xi=2e^{-2\Phi}$
and the system (\ref{XT0})-(\ref{B0}) becomes
\begin{eqnarray}
{d(\xi^2 t^2)\over d\Phi} &=& -\,\xi^2 t^2 + 2\,\xi - \Lambda\,\xi^2 
- 2Q^2 + {c_0^2\over 2}\,{\xi^2 t^2\over B^2},\label{XT1}\\
{1\over 2B^2}{dB^2\over d\Phi} &=& {2\,\xi - \Lambda\,\xi^2 - 
2Q^2 \over \xi^2 t^2}.\label{B1}
\end{eqnarray}
All the $n\neq 2$ solutions can be generated from the solutions of 
(\ref{XT1})-(\ref{B1}) for the vanishing cosmological constant $\Lambda=0$.
Indeed, in the general case let us introduce the new, rescaled, variables
and constants:
\begin{eqnarray}
\widetilde{\Phi} &=& 2\left({n-1\over n}\right)\Phi\qquad \Longrightarrow\qquad
\widetilde{\xi} = 2e^{-2\widetilde{\Phi}} = 2e^{-4\left({n-1\over n}\right)\Phi},
\label{t1}\\
\widetilde{(\xi^2 t^2)} &=& {1\over\lambda}\left({2\over n}\right)^2 (\xi^2t^2),
\qquad\qquad \widetilde{B^2} = \left({2\over n}\right)^2 B^2,
\label{t2}\\
\widetilde{Q^2} &=& {2\over n(n-1)}\,{Q^2\over\lambda},\qquad\qquad
\widetilde{c^2} = {2\over n(n-1)}\,c^2_0.\label{t3}
\end{eqnarray}
Clearly, we assume that $\lambda\neq 0$, the case $\lambda=0$ will be considered
separately. Note that the quantities with tilde are {\it not} necessarily positive
and this is an essential difference between the cases of $\lambda=\pm 1$. 
Substituting (\ref{t1})-(\ref{t3}) into (\ref{XT0})-(\ref{B0}), we get
\begin{eqnarray}
{d\widetilde{(\xi^2 t^2)}\over d\widetilde{\Phi}} &=& -\,\widetilde{(\xi^2t^2)} + 
2\,\widetilde{\xi} - 2\widetilde{Q^2} + {\widetilde{c^2}\over 2}\,
{\widetilde{(\xi^2 t^2)}\over \widetilde{B^2}},\label{XT2}\\
{1\over 2\widetilde{B^2}}{d\widetilde{B^2}\over d\widetilde{\Phi}} &=& {2\,
\widetilde{\xi} - 2\widetilde{Q^2} \over \widetilde{(\xi^2 t^2)}}.\label{B2}
\end{eqnarray}

Thus, all the solutions for the Einstein-Maxwell-scalar system in arbitrary
dimension either spherically symmetric ($\lambda=1$) or ``hyperbolically'' 
symmetric ($\lambda=-1$) is generated from the spherically symmetric solutions 
in four dimensions by reversing the transformation (\ref{t1})-(\ref{t3}).

\section{Exact solutions for arbitrary dimension and $\lambda$}\label{sol}

The general solution for the $n=2, \lambda=1$, i.e. for spherically symmetric
four-dimensional Einstein-Maxwell-scalar field equations is well known 
\cite{tekin}. In our formalism, it reads as follows:
\begin{equation}
\widetilde{\xi} = 2\,h,\qquad \widetilde{(\xi^2 t^2)}=-\,{1\over g}\,
\left({dh\over dr}\right)^2,\qquad
\widetilde{B^2} = - \,\widetilde{(\xi^2 t^2)}\,f,\label{n1sol}
\end{equation}
where the functions $f=f(r),\, g=g(r),\, h=h(r)$ are given by
\begin{equation}
f = {(x^2 - 1)^{2}\over {\cal D}^2},\qquad g ={\cal D}^2,\qquad 
h = {\cal D}^2\,r^2.\label{fgh}
\end{equation}
Here the function
\begin{equation}
{\cal D}(x):={k^2\,(x + 1)^{2\mu} - (x - 1)^{2\mu}\over (x^2 - 1)^{\mu -1}}
\end{equation}
depends on $r$ via the auxiliary variable 
\begin{equation}
x : = \,-\,{M\over r},\label{x}
\end{equation}
and $M$, $k^2$ and $\mu$ are arbitrary integration constants. The main 
properties of the function ${\cal D}(x)$ can be seen from the differential
identities it satisfies:
\begin{eqnarray}
2\,{d\over dx}\left({1\over {\cal D}}{d{\cal D}\over dx}\right) +
\left({1\over {\cal D}}{d{\cal D}\over dx}\right)^2 &=&
- \,{16\,\mu^2\,k^2\over {\cal D}^2} - \,{4(1-\mu^2)\over (x^2 - 1)^2},\label{D1}\\
\left({1\over {\cal D}}{d{\cal D}\over dx}\right)^2 +\,{4x\over x^2 - 1}\,
\left({1\over x} -\,{1\over {\cal D}}{d{\cal D}\over dx}\right) &=& 
\quad{16\,\mu^2\,k^2\over {\cal D}^2} - \,{4(1-\mu^2)\over (x^2 - 1)^2}.\label{D2}
\end{eqnarray}

Using (\ref{n1sol})-(\ref{D2}) in (\ref{XT2})-(\ref{B2}), we find explicitly the 
integration constants for a general solution with an arbitrary $n$ and $\lambda$:
\begin{eqnarray}
Q^2 &=& 16\,n(n-1)\,M^2\,\mu^2\,k^2\,\lambda,\label{Q2}\\
c_0^2 &=& 16\,n(n-1)\,M^2\,(1 -\mu^2)\,.\label{c2}
\end{eqnarray}
Substituting (\ref{n1sol}) into (\ref{eq3}), (\ref{TB}), (\ref{met1}) and (\ref{KK0}),
one obtains the metric:
\begin{equation}
g = -\,\lambda f\,d\tau^2 + {g\,h^{-\left({n-2\over n-1}\right)}\over \lambda (n-1)^2}
\,dr^2 + h^{{1\over n-1}}\,d\Omega_\lambda^2,\label{met2}
\end{equation}
where $d\Omega_\lambda^2$ is the line element on the $n$-dimensional space of 
constant curvature $\lambda$. 

It is worthwhile to note that the equation (\ref{Q2}) demands that for the spherical 
symmetry ($\lambda=1$) we take $k^2\geq 0$, whereas for the ``hyperboloidal'' symmetry
 ($\lambda=-1$) one should take $k^2\leq 0$.

The scalar field is obtained after some algebra from the first integral (\ref{Qc2}),
\begin{equation}
\phi = \sqrt{(1-\mu^2)\left({n\over n-1}\right)}\,
\log\left\vert{x-1\over x+1}\right\vert.
\end{equation}
As for the Maxwell field, the electromagnetic potential, $A =A_0\,d\tau$,
is obtained form (\ref{Qc1}) in the form:
\begin{equation}
A_0=-\,\sqrt{\lambda k^2\left({n\over n-1}\right)}\ {1\over 1-k^2}\ {(x + 1)^{2\mu}
 - (x - 1)^{2\mu}\over {k^2\,(x + 1)^{2\mu} - (x - 1)^{2\mu}}},\label{A1}
\end{equation}
for the case when $k^2\neq 1$. When $k^2=1$, we have 
\begin{equation}
A_0 = - \,\sqrt{\lambda k^2\left({n\over n-1}\right)}\,
{(x + 1)^{2\mu}\over {(x + 1)^{2\mu} - (x - 1)^{2\mu}}}.\label{A2}
\end{equation}
However, both potentials give rise to the same Maxwell field strength two-form
\begin{equation}
F = dA = 4\,M\,\mu\,\sqrt{\lambda k^2\left({n\over n-1}\right)}\,
{x^2 - 1\over h}\,d\tau\wedge dr,
\end{equation}
which is valid for all possible values of $k^2$. A straightforward check shows
that $\hbox{$\scriptstyle{\#}$}F = -\,Q\,{\buildrel (n)\over \eta}$, where 
${\buildrel (n)\over \eta}$ is the volume $n$-form of the internal space. Thus
indeed the integration constant $Q$ is the electric charge of the solution. 

For completeness, let us give the dilaton field explicitly:
\begin{equation}
\Phi = - {1\over 4}\,\left({n\over n-1}\right)\,\log h.
\end{equation}

\section{Discussion}\label{disc}

It is instructive to compare our solutions to the higher-dimensional black holes
in the Brans-Dicke-Maxwell theory. The corresponding Lagrangian can be written 
(with the help of a suitable conformal transformation and field redefinitions) as 
\begin{equation}
L = -\,{1\over 2}\,{\cal R}^{AB}\wedge\eta_{AB} - {1\over 2}e^{-b\phi}\,F\wedge
\hbox{$\scriptstyle{\#}$} F -\,{1\over 2}\,d\phi\wedge\hbox{$\scriptstyle{\#}$} 
d\phi,\label{LBD}
\end{equation}
with some constant $b$. When $b=0$, this reduces to the Lagrangian (\ref{Lhigh}).
The direct investigation shows that the condition $b\neq 0$ is crucial for the
existence of non-trivial black hole solutions: the exact configurations were found
in \cite{GM,cai3}. They, however, cannot be used for obtaining the solutions of
(\ref{Lhigh}) from that of (\ref{LBD}) by taking the limit $b\rightarrow 0$ in
these configurations. The resulting solutions have trivial (constant) scalar
field. In contrast, our solutions are characterized by the nontrivial scalar 
field, in general.

In order to analyse the physical meaning of the constant parameters $\mu, M$ and 
$k^2$, it is convenient to replace the radial coordinate $r$ by a new variable
\begin{equation}
\rho = M\left( x + {1\over x}\right).\label{rx}
\end{equation}
Then the solution is rewritten in the form:
\begin{eqnarray}
g &=& -\,\lambda\,f\,d\tau^2 + {h^{-\left({n-2\over n-1}\right)}
\over\lambda (n-1)^2f}\,d\rho^2 + h^{{1\over n-1}}\,
d\Omega_\lambda^2,\label{met7}\\
F &=& 4\,M\,\mu\,\sqrt{\lambda k^2\left({n\over n-1}\right)}\,
{1\over h}\,d\tau\wedge d\rho,\label{max3}\\
\phi &=& \sqrt{{(1-\mu^2)\over 4}\left({n\over n-1}\right)}\,
\log\left\vert{1 - y\over 1 + y}\right\vert.\label{phi3}
\end{eqnarray}
Here $f=f(y)$ and $h=h(y)$ are given by
\begin{eqnarray}
f &=& {(1 - y^2)^{\mu}\over\left[k^2 (1 + y)^\mu - 
(1 -y)^\mu\right]^2},\label{f1}\\
h &=& \rho^2\,(1 - y^2)^{1-\mu}\,
\left[k^2 (1 + y)^\mu - (1 -y)^\mu\right]^2.\label{h1}
\end{eqnarray}
in terms of the variable $y = 2M/\rho$. 

Unfortunately, when $n>2$ it is impossible to find a transformation, 
in a closed analytic form, to the Schwarzschild type coordinates. The 
above transformation (\ref{rx}) brings the metric most closely to the 
Schwarzschild form and it is suitable for the analysis of asymptotic 
behaviour of the metric. Expanding the functions (\ref{f1})-(\ref{h1}) 
and the equations (\ref{met7})-(\ref{max3}) in the asymptotic spatial 
region in powers of $y$ (i.e., $\rho^{-1}$), we find for the metric 
\begin{equation}
g\approx -\lambda\left(1 -{4M\mu(k^2 + 1)\over\widetilde{\rho}^{n-1}}\right)\,
d\widetilde{\tau}^2 +\lambda\left(1 +{4M\mu(k^2+1)\over\widetilde{\rho}^{n-1}}
\right)\,d\widetilde{\rho}^2 + \widetilde{\rho}^2\,d\Omega_\lambda^2,\label{asy1}
\end{equation}
and for the Maxwell and scalar fields
\begin{eqnarray}
F &\approx& {4M\mu\sqrt{\lambda k^2\,n(n-1)}\over \widetilde{\rho}^n}\,
d\widetilde{\tau}\wedge d\widetilde{\rho},\label{asy2}\\
\phi &\approx& -\,\sqrt{(1-\mu^2)
\left({n\over n-1}\right)}\,{2M\,(k^2-1)\over \widetilde{\rho}^{n-1}}.
\end{eqnarray}
Here we accompanied the expansion by the change of the asymptotic coordinates
\begin{equation}
\widetilde{\tau} = {\tau\over (k^2 -1)},\qquad 
\widetilde{\rho} = \left[(k^2 -1)\,\rho\right]^{1\over n-1}.\label{tiro}
\end{equation}
Recall that $\widetilde{\rho}^{1-n}$ is a fundamental
solution of the Laplace equation in a $(1+n)$-dimensional space [we consider 
the $(2+n)$-dimensional space{\it time}]. The equations (\ref{asy1})-(\ref{asy2}) 
demonstrate that total mass $m$ and electric charge $Q$ are constructed
from $\mu, M, k^2$ as follows:
\begin{equation}
m = 2M\mu\lambda(k^2 + 1),\qquad\qquad 
Q = 4M\mu\sqrt{\lambda k^2\,n(n-1)}.\label{em1}
\end{equation}
The latter relation is in complete agreement with (\ref{Q2}). Note that 
for the spherical symmetry the electric charge and the mass always satisfy 
the inequality ${Q/m}\leq\sqrt{n(n-1)}$. Equality corresponds to the extremal 
charged solution. For obvious reasons, the combination $\sigma = -\,2M\,
(k^2-1)\sqrt{(1-\mu^2)\left({n\over n-1}\right)}$ can be called a scalar 
charge of the solution. This interpretation conforms with (\ref{c2}).

As one notices, the above identifications of integration constants are only
valid when $k^2\neq 1$. Besides that, there are several other special cases 
to which our general solutions reduce for particular values of the constant 
parameters $\mu$ and $k^2$. These cases should be considered separately.
In particular, from (\ref{met7})-(\ref{h1}) it is straightforward to see that 
the choice $\mu =\pm 1$ yields the charged black hole configurations which 
extend the Reissner-Nordstrom solution to a $(2 + n)$-dimensional spacetime. 
The scalar field vanishes for these solutions. One more transformation of the 
radial coordinate, $R = h^{1\over 2(n-1)}$, or explicitly, $R = [(k^2 - 1)\rho
+ 2M(k^2 +1)^2]^{1\over n-1}$, brings the gravitational and Maxwell fields to:
\begin{eqnarray}
g &=& -\,f\,d\widetilde{\tau}^2 + {dR^2\over f} + R^2\,d\Omega_\lambda^2,
\quad {\rm with}\quad f =\lambda - {2m\over R^{n-1}} + 
{Q^2\over n(n-1)R^{2n-2}},\label{met6}\\
F &=& {Q\over R^n}\,d\widetilde{\tau}\wedge dR.
\end{eqnarray}
Here $m$ and $Q$ are given by (\ref{em1}), and $\widetilde{\tau}$ is defined 
by (\ref{tiro}). This is the higher dimensional and topological generalization
of the standard Reissner-Nordstrom solution \cite{myers,cai2}.

As we already repeatedly mentioned, the case $k^2 =1$ requires special analysis.
For $k^2 =1$ {\it and} $\mu =\pm 1$, in the case of the spherical symmetry 
($\lambda =1$), we find from (\ref{met7}) a higher dimensional generalization 
of the Bertotti-Robinson solution \cite{bertrob}. It is described by the constant 
electromagnetic field $F = (4M)^{-1}\sqrt{{n\over n-1}}\,d\tau\wedge d\rho$ and  
the metric 
\begin{equation}
g = -\,(\rho^2 -4M^2)\,d\tau^2 + (4M)^{-\left({2\over n-1}\right)}\left(
{d\rho^2\over (n-1)^2(\rho^2 -4M^2)}  + \,d\Omega_\lambda^2\right)\label{br}
\end{equation}
of a direct product of the two-dimensional de Sitter spacetime and an $n$-sphere.
Such a configuration is everywhere regular. The situation is however more 
complicated when $k^2 =1$ {\it but} $\mu \neq\pm 1$. Then the electromagnetic
field and the metric are only approaching the constant Bertotti-Robinson 
configuration (\ref{br}) in the limit of large $r\rightarrow\infty$. But both
the curvature and the electromagnetic invariants have singularities at finite
values of $\rho$ for which $h=0$. This is a direct consequence of the presence
of the nontrivial scalar field. Such solutions can be called higher dimensional
asymptotically Bertotti-Robinson spacetimes. 

Another special case which deserves being mentioned, is $\mu =0$. Then from 
(\ref{f1})-(\ref{h1}) one finds $f = (k^2 -1)^{-2}$ and $h = (k^2 -1)^2\,
\rho^2\,(1 - y^2)$ (thus one has to take $k^2\neq 1$), and the metric reduces 
to [denoting $\rho_0 = 2M(k^2 - 1)$]
\begin{equation}
g = -\,d\widetilde{\tau}^2 + {(\widetilde{\rho}^2 - 
\rho_0^2)^{-\left({n-2\over n-1}\right)}\over (n-1)^2}
\,d\widetilde{\rho}^2 + (\widetilde{\rho}^2 - 
\rho_0^2)^{{1\over n-1}}\,d\Omega_\lambda^2.\label{met8}
\end{equation}
According to the above analysis, this solution has no mass and electric charge,
see (\ref{em1}), but the scalar charge is nontrivial. Such a configuration 
represents a higher dimensional generalization of the static Roberts solution
\cite{rob}. Recently an interesting study \cite{frol} of nonstatic extensions 
of the Roberts solution revealed, in any dimension $n$, the continuously 
self-similar solutions which describe collapse of scalar field with critical 
behaviour. 

\section{Acknowledgements}

The authors are grateful to TUBITAK for the support of this research. YNO
is also grateful to the Department of Physics, Middle East Technical 
University, for the warm hospitality.




\begin{references}                    
\bibitem{highlow}
M.J. Duff, B.E.W. Nilsson, and C.N. Pope, 
{\sl Phys. Rept.} {\bf 130} (1986) 1; 
J.D. Brown, {\it Lower dimensional gravity} 
(World Scientific: Singapore, 1988) 152 p.;
S. Carlip, {\it Quantum gravity in 2+1 dimensions} (Cambridge Univ. Press:
Cambridge, 1998). 
\bibitem{manntop}
R.B. Mann, 
in: {\sl ``Proc. of Workshop on the internal structure of black holes and 
spacetime singularities'', 29 June - 3 July 1997, Haifa, Israel}. Eds. L.M. Burko 
and A. Ori (Inst. of Phys.: Bristol, 1997; Isr. Phys. Soc.: Jerusalem, 1997) 311-342. 
\bibitem{cai1}
R.-G. Cai and Y.-Z. Zhang
{\sl Phys. Rev.} {\bf D54} (1996) 4891; 
D. Klemm, 
{\sl Class. Quantum Grav.} {\bf 15} (1998) 3195; 
D. Klemm, V. Moretti, and L. Vanzo, 
{\sl Phys. Rev.} {\bf D57} (1998) 6127. 
\bibitem{lemos}
J.P.S. Lemos and V.T. Zanchin,
{\sl Phys. Rev.} {\bf D53} (1996) 4684; 
N. Kaloper, 
{\sl Phys. Rev.} {\bf D48} (1993) 4658; 
W.G. Anderson and N. Kaloper, 
{\sl Phys. Rev.} {\bf D52} (1995) 4440. 
\bibitem{cai2}
R.-G. Cai, J.-Y. Ji, and K.-S. Soh, 
{\sl Phys. Rev.} {\bf D57} (1998) 6547. 
R.-G. Cai and K.-S. Soh,
{\sl Phys. Rev.} {\bf D59} (1999) 044013;
\bibitem{jan}
A.I. Janis, E.T. Newman, and J. Winicour, 
{\sl Phys. Rev. Lett.} {\bf 20} (1968) 878. 
\bibitem{xan}
B.C. Xantopoulos and T.E. Dialynas, 
{\sl J. Math. Phys.} {\bf 33} (1992) 1463.
\bibitem{agnese}
A.G. Agnese and M. La Camera, 
{\sl Phys. Rev.} {\bf D31} (1985) 1280.
\bibitem{GM}
G.W. Gibbons and K. Maeda,
{\sl Nucl. Phys.} {\bf B298} (1988) 741. 
\bibitem{kim}
S.-W. Kim and B.H Cho, 
{\sl Phys. Rev.} {\bf D40} (1989) 4028. 
\bibitem{youm}
D. Youm, 
{\sl Phys. Rept.} {\bf 316} (1999) 1. 
\bibitem{cai3}
R.-G. Cai and Y. S. Myung, 
{\sl Phys. Rev.} {\bf D56} (1997) 3466. 
\bibitem{myers}
R.C. Myers and M.J. Perry, 
{\sl Ann. Phys. (NY)} {\bf 172} (1986) 304.
\bibitem{PR} 
F.W. Hehl, J.D. McCrea, E.W.  Mielke, and Y. Ne'eman,
{\sl Phys. Rep.} {\bf 258} (1995) 1.
\bibitem{2drc}
E.W. Mielke, F. Gronwald, Yu.N. Obukhov, R. Tresguerres, and F.W. Hehl,
{\sl Phys. Rev.} {\bf D48} (1993) 3648; 
Yu.N. Obukhov, 
{\sl Phys. Rev.} {\bf D50} (1994) 5072; 
Yu.N. Obukhov, S.N. Solodukhin, and E.W. Mielke, 
{\sl Class. Quantum Grav.} {\bf 11} (1994) 3069. 
\bibitem{tworev}
Yu.N. Obukhov and F.W. Hehl, {\it Black holes in two dimensions}, in: 
{\sl Proceedings of 179.WE-Heraeus Seminar ``Black Holes Theory and 
Observations'', Bad Honnef, Germany, 18-20 Aug 1997},  Eds. F.W. Hehl, 
C. Kiefer, and R.J.K. Metzler (Springer: Berlin) {\sl Lect. Notes Phys.}
{\bf 514} (1998) 289. 
\bibitem{tekin}
T. Dereli, 
{\sl Phys. Lett.} {\bf B161} (1985) 307;
T. Dereli and A. Eri\c{s}, 
{\sl Do\v{g}a} {\bf A9} (1985) 30; 
A. Eri\c{s}, 
{\sl J. Math. Phys.} {\bf 18} (1976) 824.
\bibitem{bertrob}
B. Bertotti, 
{\sl Phys. Rev.} {\bf 116} (1959) 1331; 
I. Robinson, 
{\sl Bull. Acad. Pol. Sci.} {\bf 7} (1959) 351. 
\bibitem{rob}
M.D. Roberts, 
{\sl Gen. Rel. Grav.} {\bf 21} (1989) 907; 
M.D. Roberts, 
{\sl J. Math. Phys.} {\bf 37} (1996) 4557. 
\bibitem{frol}
A.V. Frolov, 
{\sl Class. Quantum Grav.} {\bf 16} (1999) 407. 
\bibitem{CQG} T. Dereli, Yu. N. Obukhov, {\it On the universality of low-energy string model}\\
To appear in {\sl Class.Quantum Grav.} (gr-qc/9909045)
\end{references}
\end{document}